\documentclass[traditabstract]{aa} 
\usepackage{graphicx}
\usepackage{txfonts}
\begin{document}
   \title{Enhanced emission from GRB 110328A\\could be evidence for tidal disruption of a star}


   \author{U. Barres de Almeida\thanks{Corresponding Author; E-mail \email{ulisses@mppmu.mpg.de}}
          \and
          A. De Angelis\thanks{\email{alessandro.de.angelis@cern.ch}. Also at INFN and INAF Trieste; on leave of absence from University of Udine, Italy.}
          }

   \institute{Max-Planck-Institut f\"ur Physik
(Werner-Heisenberg-Institut),
M\"unchen, Germany.}             

   \date{ }
   
     \abstract{On March 28, Swift's Burst Alert Telescope discovered a source in the constellation Draco when it erupted in a series 
of X-ray blasts. The explosion, catalogued as gamma-ray burst (GRB) 110328A, repeatedly flared in the following days, 
making the interpretation of the event as a GRB unlikely. Here we suggest that the event could be due to the tidal 
disruption of a star that approaches the pericentric distance of a black hole, and we use this fact to derive bounds on the 
physical characteristics of such system, based on the variability timescales and energetics of the observed X-ray emission.
}
   
   
                  \titlerunning{Evidence for tidal disruption of a star by a black hole}

\maketitle


%

\section{Introduction}

The capture and tidal disruption of a star in the vicinity of a super-massive black hole (SMBH) has been considered as one
of the ways in which an active galactic nucleus (AGN) can be fueled (\cite{rees1}). In particular, this phenomenon has been
used by several authors to support the existence of ``dormant quasars'', consistent with 
the now widely-believed conjecture that a SMBH is harboured at the nucleus of every large galaxy. Despite much detailed 
theoretical work in the field, direct observational evidence for the tidal disruption and accretion of a star by a SMBH 
is somewhat scarce. Studies based on soft X-ray survey data, for example, have found evidence for intense and short-lived 
flare periods in otherwise low-active galaxies that are consistent with the theoretical prognostics for such an event 
(e.g., \cite{komossa99}, \cite{komossa04}, \cite{indirectev}). Although the timescales associated with this phenomenon are such that it would be possible to 
follow its unfolding in time with great detail, such detection had not been reported yet.
\\
\\
On March 28, the Burst Alert Telescope (BAT) onboard the Swift satellite triggered on a source in the constellation of 
Draco when it erupted in a series of X-ray blasts. The satellite determined the position for the explosion, initially 
classified as gamma-ray burst (GRB) 110328A (a.k.a. Swift J164449.3+573451), at coordinates $RA=251.2054$ h and 
$DEC=57.5808$ degrees, at a location in the sky where no catalogued X-ray source existed (\cite{GCN 11823}). 
GRB 110328A repeatedly flared in the days following its discovery, presenting a behaviour incompatible with that of a 
gamma-ray burst\footnote{Even if this is the case, for simplicity we will continue to call the source by its originally 
catalogued name, GRB110328A.}(\cite{GCN 11824}). Identification of an optical counterpart from pre-burst images
of the region followed shortly, revealing an object of magnitude $R \sim 22$ (\cite{GCN 11827}) which was independently
confirmed by others (\cite{GCN 11830}). The extragalactic nature of the optical counterpart was 
determined from observations of H$\beta$ and OIII emission lines at a common redshift of $z \sim 0.35$ by the Gemini 
Telescope (\cite{GCN 11833}).

Figure 1 shows the brightness changes recorded by Swift in the first two weeks after the trigger\footnote{Swift data 
obtained directly from the Swift/UK mirror website \textit{www.swift.ac.uk} processed following standard analysis 
described in (\cite{Evans}).}. The light-curve clearly shows the 
highest activity period concentrated in the first 2 days of observations, with the two brightest flare peaks separated by 
approximately one day and followed by smaller peaks, of decaying relative amplitude and durations of the order of $10^4$ 
seconds. The emission then attains a minimum and after a few days regains intensity, but now at a more stable rate and 
with an average intensity about one order of magnitude fainter than that of the bright flares. As of today, this emission 
persists.    

\begin{figure}
\begin{center}
\resizebox{\columnwidth}{!}{\includegraphics{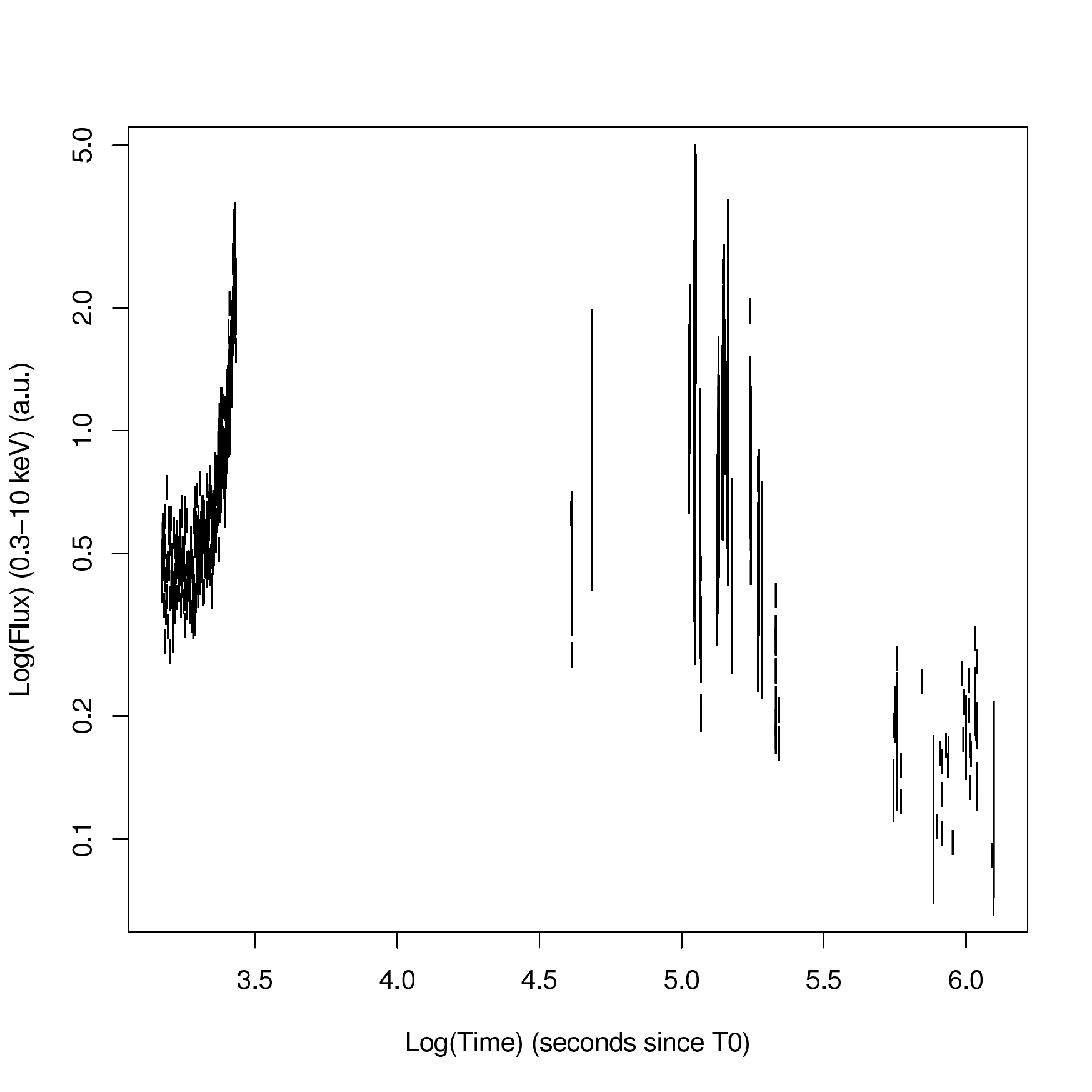} }
\end{center}
\caption{Swift XRT light-curve of GRB 110328A (a.k.a. Swift J164449.3+573451). The data was directly obtained from 
the Swift website (\cite{Evans}). The flux was normalised to the mean value of the first two nights and is given in
these arbitrary units.}
\label{fig:1}       
\end{figure}

In this note we will discuss the basic evidence in favour of the X-ray data of GRB 110328A being the direct 
observational signature of a star being disrupted and subsequently accreted after a close passage to a SMBH in the 
centre of the galaxy identified as the probable optical counterpart of the X-ray source.
  

\section{Dynamical considerations and energetics}


The optical counterpart of GRB 110328A was identified with a red galaxy at $z \sim 0.35$. At this distance, the absolute
R-band magnitude of the host is of about $M_{\rm{R}} \simeq -18$. According to the mass-luminosity relation for spheroids, this magnitude 
implies a mass for the SMBH at the centre of the galaxy of the order of $M_6 \simeq 10$ (\cite{Graham}), which will dictate 
the dynamical scales of our system. In our notation $M_6$ denotes the black hole mass in units of $10^6$ $M_{\odot}$. Observe that 
according to \cite{Graham}, $10^6$ and $10^8$ M$_{\odot}$ can be considered as strong lower and upper bounds for the mass of the SMBH, respectively.

The distinctive observational signature of the capture of a star by a SMBH (\cite{rees1}) is a series of more or less 
short-lived transient periods of outburst which happen when the star is tidally disrupted by a close passage to 
the central object, and part of the stripped material (or debris) accretes  the central object. For very close orbits
$a \gtrsim R_{\rm{T}}$ -- where $a$ is the semi-major axis of the orbit and $R_{\rm{T}}$ is the  tidal radius of the black hole -- the 
star is not expected to survive many passages, and could be entirely destroyed in a single flyby. Furthermore, after the 
disruption not all the material is expected to stay in a bound orbit about the hole, but up to half can be 
expelled from the system in hyperbolic orbits at high speeds. 

According to Rees (1988), the luminosity at the peak of the infall rate, corresponding to the times immediately after 
each pericentre passage, and associated to the accretion of the most tightly bound debris from the disruption, can be of 
the order or higher than the Eddington luminosity
\begin{equation} 
L_{Edd}\simeq 1.25\times10^{44}M_6 ~~~\rm{erg} \cdot \rm{s}^{-1} ,  
\end{equation}
which in our case is $L_{Edd}\sim 1.25 \times 10^{45}$ erg$\cdot$s$^{-1}$. The peak luminosity of the event as inferred 
from the X-ray observations was of the order of $5\times 10^{48}$ erg$\cdot$s$^{-1}$ (\cite{GCN 11843}), and the 
average luminosity for the first day of emission (its brightest period, with duration $\sim 10^5$ s) was approximately 
$L_X \sim 2.5\times10^{47}$erg$\cdot$s$^{-1}$ (\cite{GCN 11847}), about 2 orders of magnitude above $L_{Edd}$ for
$M_6 \sim 10$. 
Although accretion at the initial phases can in principle proceed at super-Eddington rates (\cite{rees1}), this large 
discrepancy between $L_X$ and $L_{Edd}$ can be taken as evidence for beaming of the emission by an unknown $\Gamma$ factor.
In order to recover the more comfortable value of $\sim L_{Edd}$ from the observed X-ray luminosities, we need therefore to
introduce the factor
\begin{equation} \label{beaming}
\Gamma^2M_6 = \frac{L_X}{L_{Edd}} \simeq  10^3,  
\end{equation}  
which for $M_6=10$ implies a beaming of $\Gamma \sim 10$.

This beaming could be realised in the form of jets, as usually observed from quasars (see Figure 2). If the 
observed emission has origin in jets then it is likely to be at least in part nonthermal, though contributions from a
thermal emission from the accretion disk might be present. 


\section{Characteristics of the tidal disruption event}


A star of mass $m_*$ and radius $r_*$ becomes vulnerable to tidal distortions (\cite{rees1}) when the pericentre  
of its orbit becomes small enough to be comparable to the tidal radius $R_{\rm{T}}$. In terms of the Schawrzschild radius 
$R_S \simeq 3\times 10^{11} M_6$ cm, this quantity is:
\begin{equation} 
\left(\frac{R_T}{R_{S}}\right) \simeq  15 M_6^{-2/3} \left(\frac{r_*}{r_{\odot}}\right) \left(\frac{m_*}{m_{\odot}}\right)^{-1/3} > 1,  
\label{eq:ratio}
\end{equation}
where the condition of $R_T/R_S > 1$ corresponds\footnote{If $R_{S}\sim R_{\rm{T}}$ then the 
tidal radius would fall inside the Schwarzschild radius of the black hole (BH) and the Newtonian approximation adopted here
would no longer be valid (\cite{rees2}).} to $M_6 \lesssim 100$.  The tidal disruption process as envisaged here is 
depicted in Figure 2.

\begin{figure*}
\begin{center}
\resizebox{1.5\columnwidth}{!}{\includegraphics{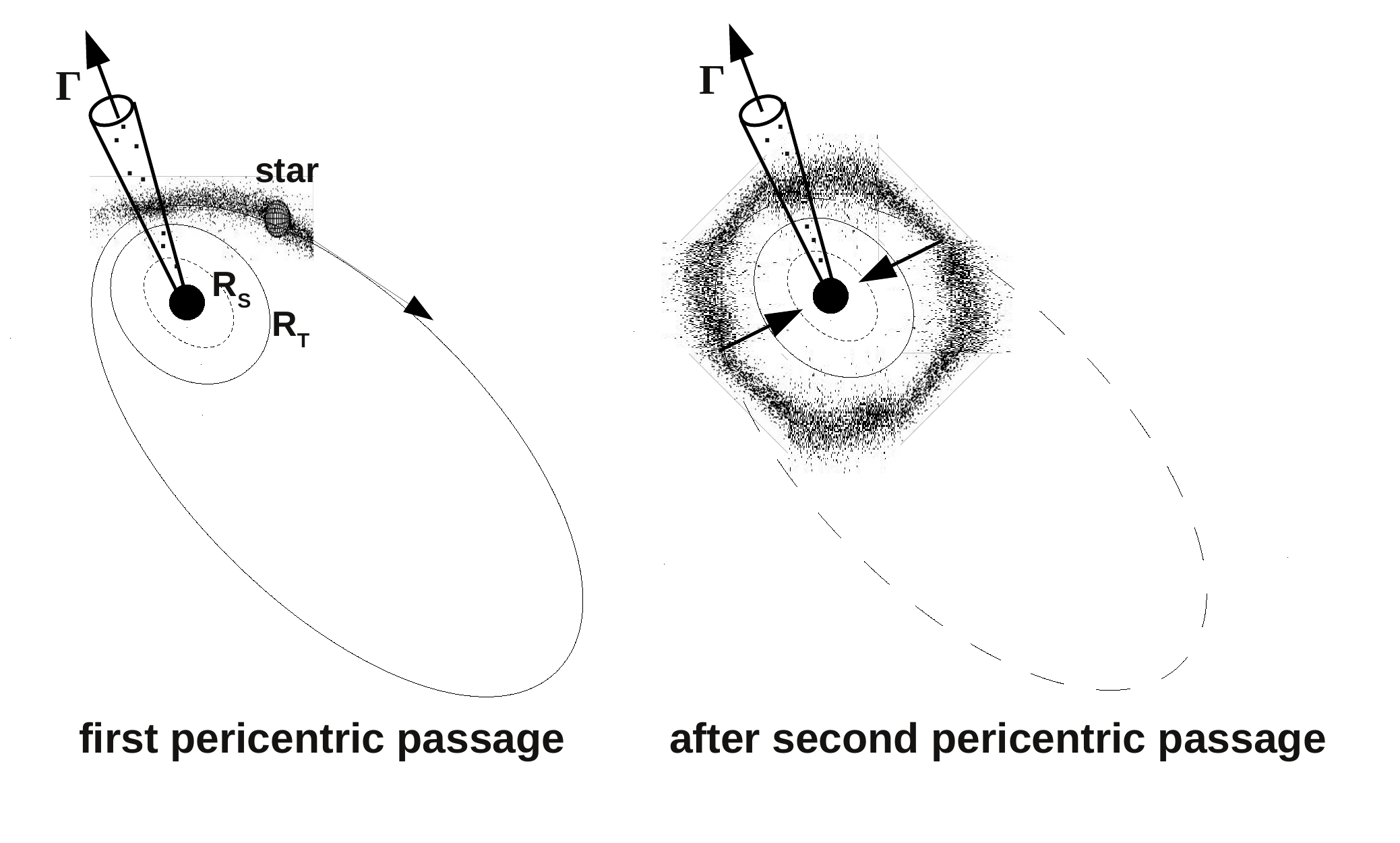} }
\end{center}
\caption{A solar-type star approaching a massive black hole on a elliptic orbit with pericentre distance $R_T$ is 
distorted and spun up during, and then tidally disrupted, leaving behind a disk of debris that are accreted by the hole.}
\label{fig:2}       
\end{figure*}

Let us therefore suppose that a star with an orbit of semi-major axis $a$ will have a periastron passage by the SMBH at a distance
$\sim R_{\rm{T}}$. The  period of the orbital motion for such a star  is: 
\begin{equation}
T \simeq 88 M_6 \left(\frac{a}{R_S}\right)^{3/2} \, \rm{s}.
\label{eq:period}
\end{equation}

The X-ray light-curve observed for GRB 110328A presents two distinctively bright peaks separated by approximately 
10$^5$ s, which we can tentatively associate with two sequential periastron passages of the star, after which it is
completely disrupted. The bright peaks seen in the X-ray light-curve at these times are associated to the fact that
the most tighly bound debris after each passage will quickly fall inside the black hole, sustaining a strong luminosity
comparable to or higher than $L_{Edd}$ (\cite{rees1}).  

Referring to Equations \ref{eq:ratio} and \ref{eq:period}, one can see that the condition $T \simeq 10^5 
\rm{s}$ is easily verified for a SMBH close to $10^7$ $M_\odot$ when 
\begin{equation} 
\left(\frac{r_*}{r_{\odot}}\right)\left(\frac{a}{R_{\rm{T}}}\right) \sim 10.
\end{equation}
Given that $a \gtrsim R_{\rm{T}}$, this implies a value of $r_* \lesssim 10 R_\odot$, comfortably within the range for 
giant stars evolved from the main sequence (i.e., $0.5 M_\odot < m_* < 5 M_\odot$), which are likely to 
populate red ellipticals.

While the timescale between the two major flares could be related to the orbital period of the star, the duration of the 
bright flaring periods that follow them are likely to be related to the infall time of the debris in the innermost orbits. According to Rees (1988), such debris will enter in an orbit around the BH with pericentric distance $\simeq R_{T}$, and 
so the falling time is
\begin{equation} 
\tau \simeq 0.03 \, M_6^{1/2} \rm{years} \, .
\end{equation}
For a BH of $\sim 10^7 M_\odot$  this gives $\tau \sim 3\times10^6$ s, consistent with the duration  
of the high luminosity emission seen in the XRT light-curve at the beginning of the event. These timescales are of the 
same order as those for which a BH of mass $M_6 = 10$ can sustain an accretion luminosity $\sim L_{Edd}$ before
$\dot{M}$ starts to fall. This provides an independent confirmation of the order of magnitude of the BH mass. Furthermore,
it should be observed that, by the same argument (\cite{rees1}), the maximum rate at which the BH can accrete at the 
Eddington level with a radiative efficiency $\epsilon$ is
\begin{equation} 
\dot{M} \simeq  \left(0.02 \, \epsilon_{0.1}^{-1} \, M_6\right) \, M_\odot/\rm{year} \, ,
\end{equation}
corresponding, for a typical efficiency of $10\%$, to about 0.1$M_\odot$ in the observed timescale of two days, as reported in the circular GCN 11847 \cite{GCN 11847}.

Finally, we had already observed that after the second day, the X-ray emission decreased to a more or less steady flux,
a factor $\lesssim 10$ below $L_{Edd}$. Above, we had assumed that the duration of the high-luminosity state was dictated
by the infall time of the innermost bound debris from the disrupted star. Another prediction of the models of tidal
disruption is that an extended disk of material will be created, as schematically shown in Figure 2  
(\cite{gurza}). If this is the case, after the innermost bound material has been swallowed, the remaining mass will 
continue to accrete the BH but at a slower rate, dictated by the viscous timescale of the disk (\cite{gurza} and 
\cite{lynden-bell}). In our scenario, this is the likely origin of the steadier emission seem from the object during the 
last 10 days or so.

The viscous timescale of the disk has an upper bound on the free-fall timescale, and is likely to be only a modest
multiple of this value (\cite{rees1}). Therefore, if the outermost bound material is within a few $10R_{\rm{T}}$ 
of the hole, the expected timescale for accretion will be within a few months, which corresponds to a prediction that the 
X-ray flux should decrease significantly within this time after the brightest period.

\section{Conclusions}


The energetics and temporal behavior 
of the transient GRB 110328A can be explained as the first detection ``in the act'' of the disruption of a solar-type star by a 
supermassive black hole of the order of $10^7$ $M\odot$. 

The emission appears to be beamed and is likely to be at least in part nonthermal, which could imply the existence of a yet undetected higher energy component of the spectrum.

If our predictions are correct, the X-ray emission is expected to fade within a few months.

\begin{acknowledgements}
     We acknowledge M. Teshima, R. Turolla, M. Pimenta, P. Colin,  D. Paneque, G. Ghisellini for useful discussions and  comments.

\end{acknowledgements}

\end{document}